\documentclass[fleqn,usenatbib]{mnras}

\usepackage[T1]{fontenc}
\usepackage{ae,aecompl}

\usepackage{graphicx}	
\usepackage{amsmath}	
\usepackage{amssymb}	
\usepackage{amsfonts}
\usepackage{hyperref}
\usepackage{cleveref}
\crefname{section}{§}{§§}
\Crefname{section}{§}{§§}
\usepackage{longtable}
\usepackage{natbib}
\usepackage{verbatim}
\usepackage{lscape}
\usepackage{graphics}
\usepackage{epsfig}
\usepackage{multirow}
\usepackage{setspace}
\usepackage{bigdelim}
\usepackage{bigstrut}
\usepackage{floatflt}	
\usepackage{wrapfig}
\usepackage{multicol}

\usepackage{indentfirst}

\usepackage{caption}
\usepackage{subcaption}
\usepackage{verbatim}  
\usepackage{float}
\usepackage{xcolor}     

\sffamily
\def\h1{H\,{\sc i}}
\def\hh{H$_2$}

\def\ga{\raisebox{-0.3ex}{\mbox{$\stackrel{>}{_\sim} \,$}}}
\def\la{\raisebox{-0.3ex}{\mbox{$\stackrel{<}{_\sim} \,$}}}

\def\c1{C\,{\sc i}}

\def\NH3{NH$_{3}$}
\def\ch3cn{CH$_{3}$CN}

\def\kms{km s$^{-1}$}
\def\apj{ApJ}
\def\apjs{ApJSS}
\def\apjl{ApJL}
\def\aa{A\&A}
\def\aap{A\&A}

\def\aj{AJ}

\def\mnras{MNRAS}

\defcitealias{obreschkow14}{OG14}
\defcitealias{obreschkow16}{O16}

\title[AM regulates HI content in the discs of spirals]{ 
Angular momentum regulates \h1 gas content and \h1 central hole size in the discs of spirals
}

\author[C. Murugeshan et al.]{Chandrashekar Murugeshan,$^{1}$\thanks{E-mail:cmurugeshan@swin.edu.au}
Virginia Kilborn,$^{1}$ Danail Obreschkow,$^{2}$ \newauthor Karl Glazebrook,$^{1}$  Katharina Lutz,$^{1,3}$ Robert D\v{z}u{d}\v{z}ar,$^{1}$ Helga D\'{e}nes$^{4,5,6}$ \\
$^{1}$Centre for Astrophysics and Supercomputing, Swinburne University of Technology, Hawthorn, Victoria 3122, Australia \\
$^{2}$International Centre for Radio Astronomy Research, The University of Western Australia, Crawley, WA 6009, Australia \\
$^{3}$Observatoire Astronomique de Strasbourg, Universit\'{e} de Strasbourg, CNRS, UMR 7550, 67000 Strasbourg, France \\
$^{4}$Australia Telescope National Facility, CSIRO Astronomy and Space Science, P.O. Box 76, Epping, NSW 1710, Australia \\
$^{5}$Research School of Astronomy and Astrophysics, Australian National University, Canberra, ACT 2611, Australia \\
$^{6}$ASTRON, The Netherlands Institute for Radio Astronomy, Postbus 2, NL-7990 AA Dwingeloo, the Netherlands
}

\date{Accepted XXX. Received YYY; in original form ZZZ}

\pubyear{2018}
\begin{document}
\label{firstpage}
\pagerange{\pageref{firstpage}--\pageref{lastpage}}
\maketitle
\begin{abstract}
The neutral atomic hydrogen (\h1) content of spiral galaxies has been observed to vary with environment, with more \h1-deficient spirals residing in high density environments. This can be attributed to environmental effects such as ram pressure stripping and tidal interactions, which remove \h1 from the discs of galaxies. However, some spirals in low-density environments have also been observed to have relatively low \h1 mass fractions. The low densities of the Intra Galactic Medium and lack of nearby galaxies in such environments make ram pressure stripping and tidal interactions unlikely candidates of gas removal. What then could be making these spirals \h1 deficient? Obreschkow et al. introduced a parameter-free model for the neutral atomic gas fraction ($f_{atm}$), in a symmetric equilibrium disc as a function of the global atomic stability parameter ($q$), which depends on specific angular momentum. In order to examine if this model accounts for \h1-deficient galaxies in low-density environments, we have used the $M_{\textrm{\h1}} ~-$ M$_{R}$ scaling relation to select six \h1-deficient spiral galaxies and observed them with the ATCA. By measuring their $f_{\small atm}$ and $q$ values we find that the galaxies owe their observed \h1 deficiencies to low specific angular momenta. Additionally, we also find that the central \h1 hole sizes of our sample galaxies are related to their $q$ values, following the prediction of Obreschkow et al. This result brings to light the importance of angular momentum in understanding the physics of the interstellar medium in the discs of galaxies and consequently their evolution.
\end{abstract}



\section{Introduction}
\label{sec:intro}

\h1 studies of galaxies are critical to understanding how galaxies evolve, as hydrogen is the primary fuel for star formation. A number of parameters, both internal and external, are believed to play important roles in regulating the \h1 gas in galaxies and therefore their evolution. Of particular interest is the nature versus nurture debate relating to how galaxies evolve. The former supports the idea that galaxies with different morphologies tend to evolve due to fundamental differences in their internal physics, while the latter features the idea that environmental parameters affect galaxies, leading to their different evolutionary paths. The \h1 discs of galaxies being sensitive to both internal and external agents can serve as an incredible tool to substantiate the degree of influence that internal and environmental parameters can have on a galaxy. \\
\indent To aid us in assessing the \h1 content in galaxies are a number of scaling relations. The \h1 mass ($M_{\textrm{\h1}}$) and \h1 mass fraction ($M_{\textrm{\h1}}/M_{*}$) are found to correlate well (albeit with large scatter) with a galaxy's physical properties such as stellar mass, stellar surface mass density, NUV - \textit{r} colour, $R$-band magnitude, as well as other photometric magnitudes (Catinella et al.~\citeyear{catinella10},~\citeyear{catinella12},~\citeyear{catinella18}; D\'{e}nes et al.~\citeyear{denes14}; Maddox et al.~\citeyear{maddox15}; Brown et al.~\citeyear{brown15},~\citeyear{brown17}). These scaling relations serve as important tools to understand the general trend in the \h1 content among galaxies with respect to other observable properties. The origin of scatter in the scaling relations could be due to a number of reasons which may be linked to internal properties of the galaxies such as their morphology as well as environmental influences, that may affect their \h1 gas content (Cortese et al.~\citeyear{cortese11}). Identifying obvious outliers (i.e. galaxies that either have excess \h1 or significantly lower \h1 content for their given stellar mass, or colour etc) in the scaling relations and studying them, will allow us to identify and understand the reason behind their offset from the relations. In this work, we focus on galaxies that fall in the \h1-deficient part of the scaling relations.
\h1 deficiency in galaxies has originally been attributed to environmental influences. There is observational evidence that gas-deficient early-type galaxies reside in denser environments such as clusters and compact groups, while the more gas-rich late-type galaxies reside in relatively less dense environments (the morphology density relation; Dressler~\citeyear{dressler80}; Fasano et al.~\citeyear{fasano2000}). There is also evidence that the gas fraction in spiral galaxies in denser environments is relatively less compared to spirals of similar type and stellar mass in the field (Davies \& Lewis~\citeyear{davies73}; Giovanelli \& Haynes~\citeyear{giovanelli85}; Solanes et al.~\citeyear{solanes01}). This has been linked to gas removal mechanisms in dense environments such as ram pressure and tidal stripping (Gunn \& Gott~\citeyear{gunn72}; Fasano et al.~\citeyear{fasano2000}; Bekki et al.~\citeyear{bekki11}).  However, a few studies, notably Kilborn et al. (\citeyear{kilborn05}), Sengupta \& Balasubramanyam (\citeyear{sengupta06}), Hess \& Wilcots (\citeyear{hess13}) and D\'{e}nes et al. (\citeyear{denes16}) have  shown  that  some  galaxies  in  lower  density  environments such as loose groups are also HI-deficient. The low  densities  of  the  Intra  Group  and  Intra  Galactic Medium (IGM) in such environments make ram pressure and tidal stripping unlikely candidates for gas removal. What then could be driving the HI deficiency in these galaxies? \\
\indent In the last decade or so, studies of galactic baryonic angular momentum have been gaining popularity.
Angular momentum is regarded as one of the fundamental properties of galaxies. Many previous works have linked stellar angular momentum to other fundamental properties of a galaxy, such as its stellar mass and bulge-to-total ratio (Fall~\citeyear{fall83}; Romanowsky \& Fall~\citeyear{romanowsky12}). However, only recently a relation between angular momentum and atomic gas fraction in galaxies has been established. Obreschkow et al. (\citeyear{obreschkow16})[hereafter \citetalias{obreschkow16}] introduced a parameter-free quantitative model connecting the neutral atomic mass fractions of isolated disc galaxies to their specific angular momentum. A strong test for this model are disc galaxies that are particularly gas-excess or gas-deficient for their stellar content. \\
\indent Lutz et al. (\citeyear{lutz17}) studied a sample of 13
\h1-excess galaxies (with a median $\log M_{\textrm{\h1}}$[M$_{\odot}] \sim 10.4$) from the \h1 eXtreme (\h1X) survey and find that the \h1X galaxies have a low star forming efficiency owing to their large angular momenta. D\'{e}nes et al. (\citeyear{denes16}) examined 6 \h1-deficient galaxies from lower density environments such as loose groups and outskirts of clusters and find that both ram pressure stripping and tidal interactions still play important roles. Unfortunately, their resolutions did not permit them to make kinematic fits to derive the angular momenta for their sample galaxies. \\

\indent In this study we have selected 6 \h1-deficient spirals from the $M_{\textrm{\h1}} ~-$ M$_{R}$ scaling relation (see D\'{e}nes et al.~\citeyear{denes14}), where $M_{\textrm{\h1}}$ is the \h1 mass and M$_{R}$ is the \textit{R}-band magnitude, and obtained high resolution \h1 imaging from the Australia Telescope Compact Array (ATCA). The main aim of this paper is to study if specific angular momentum plays a role in regulating the atomic gas fractions among spirals in low-density environments. If confirmed, the \h1-deficiency in these galaxies and hence their offset from the scaling relations, may be explained by their low specific angular momentum. We also discuss the presence of \h1 holes in the centers of our sample galaxies. These are regions of the disc, where stellar and molecular content dominate over atomic gas, thereby leaving an \h1 depression or hole.
The theoretical models in \citetalias{obreschkow16} predict the existence of such \h1 holes and their sizes to correlate with their specific angular momentum. By testing this additional prediction, we further solidify the importance of angular momentum in influencing star formation and atomic gas regulation in the discs of galaxies.

This paper is divided into the following sections. In Section~\ref{sec:data} we describe the galaxy sample and their environments, details of the observations, as well as the methods involved in reducing the data. We also describe fitting routines, where 3D tilted ring models are fit to the spectral data cubes to derive the kinematics. The association between specific angular momentum and \h1 gas content in disc type galaxies is introduced in Section~\ref{sec:angular-mom}. In Section~\ref{sec:results} we discuss the main results of the study and conclude in Section~\ref{sec:Discussion}.


\section{Observations and data reduction}
\label{sec:data}

\begin{table*}
    \caption{The galaxy sample. Galactocentric distances have been taken from NED. V$_{\textrm{sys}}$ is the systemic velocity and R$_{\textrm{\h1}}$ is the radius at which the \h1 surface density drops to 1 M$_{\odot}$ pc$^{-2}$.}
    \label{tab:sample}
    \begin{tabular}{lccccccc}       
    \hline \hline
        Name & RA & DEC & D & V$_{\textrm{sys}}$ & $\log(M_{\star})$ & $\log(M_{\textrm{\h1}})$ & R$_{\textrm{\h1}}$ \\
             & [J2000] & [J2000] & [Mpc] & [\kms] & [M$_{\odot}$] & [M$_{\odot}$] & [kpc] \\
    \hline         
        NGC1350 & 03:31:08.10 & -33:37:43 & 24.5 & 1905 & 10.87 & 9.3 & 19.72 \\
        NGC1543 & 04:12:43.20 & -57:44:17 & 13.7 & 1176 & 10.32 & 8.75 & 15.74 \\
        NGC1792 & 05:05:14.40 & -37:58:51 & 14.2 & 1211 & 10.41 & 9.20 & 11.29 \\
        NGC2369 & 07:16:37.70 & -62:20:37 & 41.4 & 3240 & 10.80 & 9.78 & 18.47 \\
        NGC6300 & 17:16:59.50 & -62:49:14 & 13.7 & 1109 & 10.44 & 9.08 & 14.48 \\
        UGCA168 & 09:33:21.50 & -33:02:01 & 9.6 & 926 & 9.21 & 8.97 & 8.47 \\
    \hline
    \end{tabular}
    
\end{table*}
\subsection{The sample}
\label{subsec:sample}

The current study extends the investigation of Lutz et al. (\citeyear{lutz17}), where they show that angular momentum regulates the \h1 gas in \h1-excess galaxies. In this study we intend to complete the picture by studying if angular momentum also regulates the \h1 gas in \h1-deficient systems.

\begin{figure}
\hspace*{-0.3cm}
\includegraphics[width=8cm,height=6.2cm]{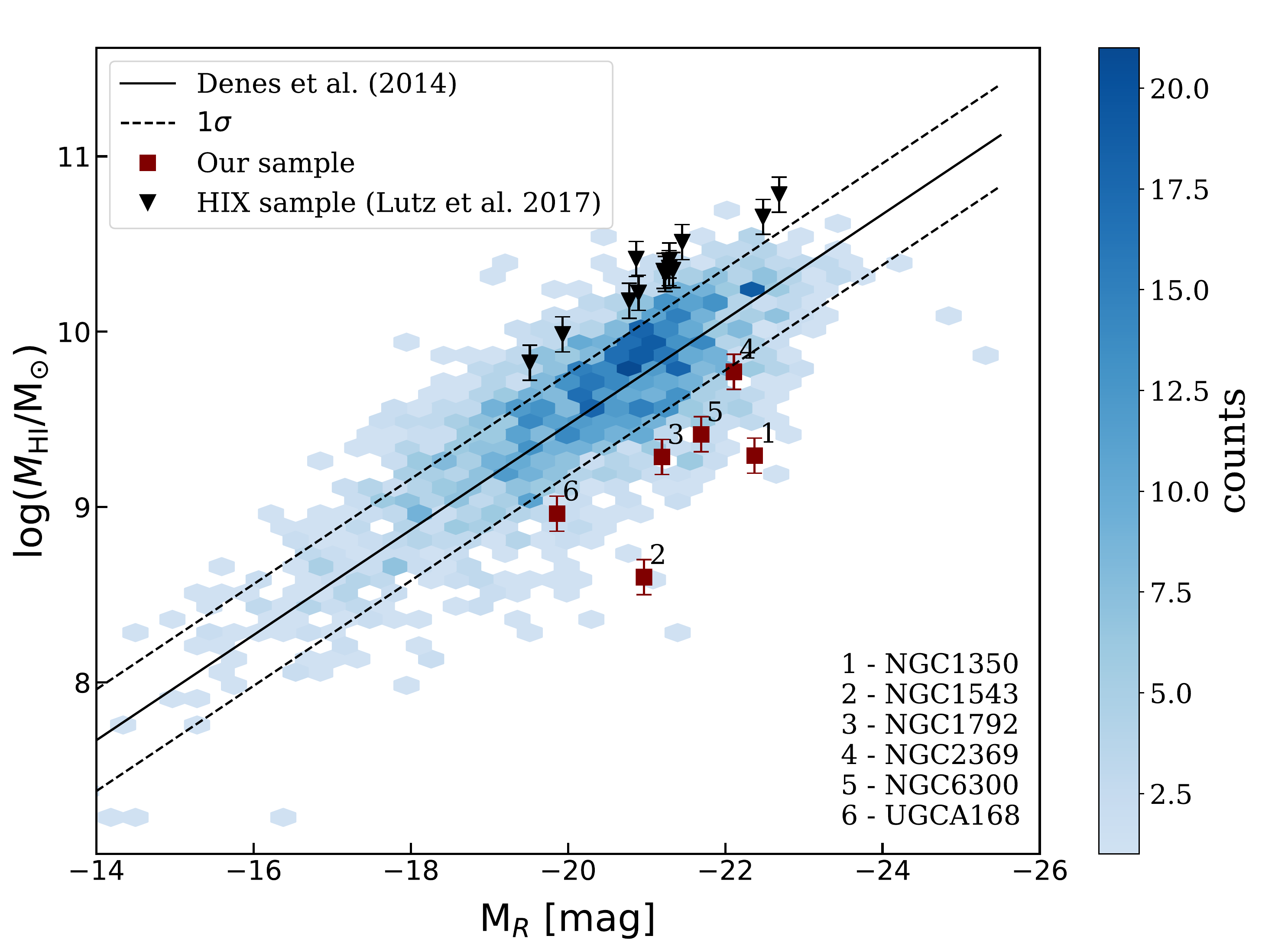}
\caption{The $M_{\textrm{\h1}} ~-$ M$_{R}$ scaling relation from D\'{e}nes  et al. (\citeyear{denes14}) for over 1700 HIPASS galaxies from the HOPCAT catalogue. The colourbar shows the number density of galaxies within each hexagon in the plot. The maroon squares represent the current sample of \h1-deficient galaxies. In contrast, also shown in black (inverted triangles) are the \h1X sample of \h1-excess galaxies from Lutz et al. (\citeyear{lutz17}).}
\label{fig:sample}
\end{figure}

We have selected six galaxies from the HIPASS Optical counterparts catalogue (HOPCAT; Doyle et al.~\citeyear{doyle05}) based on their \h1 mass, which is at least over three times less compared to average spirals for their given \textit{R}-band magnitudes. The \textit{R}-band magnitudes for the HOPCAT sample have been re-calibrated by D\'{e}nes et al. (\citeyear{denes14}).
The six \h1-deficient galaxies are plotted in the $M_{\textrm{\h1}} ~-$ M$_{R}$ scaling relation shown in Fig.~\ref{fig:sample}, where they all lie outside of the lower 1$\sigma$ cut-off in the relation. The galaxies were also required to have an isophotal diameter $D_{\textrm{25}} > 210''$, so that they may be well resolved with approximately 7-8 ATCA synthesized beam elements. This will guarantee reliable kinematic fits to the data cubes. And lastly, the galaxies were selected from low density environments. Five of the  six \h1-deficient galaxies in our sample lie at the far edges of loose groups and one galaxy lies at the outskirts of the Fornax cluster (for more details see Appendix~\ref{App:Env}). The galaxy sample with their relevant details are listed in Table.~\ref{tab:sample}. In the following subsection, we discuss briefly the environments of the sample galaxies.

\subsection{Environments}
\label{subsec:env}
In order to parameterise the local densities of the environments, we perform the standard 5th Nearest-Neighbour projected density estimates (see for example Baldry et al.~\citeyear{baldry06}). Originally, the nth Nearest-Neighbour density estimate was introduced by Dressler (\citeyear{dressler80}) while studying the local densities of groups and clusters. To compute the Nearest-Neighbour densities for our sample, we use the 6dF Galaxy Survey catalogue (6dFGS; Jones et al.~\citeyear{6dF04}). The 6dF survey is complete to $r_{F} \leq 12.75$. Since the selection criteria of galaxies for the 6dF survey is based on their \textit{K}-band magnitudes, it is designed to pick up not only bright blue (star forming) galaxies but also the older populations. A catalogue compiled from a single wide survey such as the 6dF guarantees homogeneity and uniformity, which is important for the local density estimates. To estimate the 5th Nearest-Neighbour densities, we first make a velocity cut of $\pm1000$ \kms centred on the target galaxy in order to avoid the inclusion of any foreground and/or background sources. The projected distance to the 5th nearest neighbour is then used to estimate the surface density of the local environment surrounding the target galaxy, given by $\Sigma_{5} = 5/\pi \cdot d_{5}^2$ in units of Mpc$^{-2}$. The 5th Nearest-Neighbour density estimates for our sample galaxies are listed in Table.~\ref{tab:densities}. For comparison, also shown are the density estimates for a few well known clusters and groups. The numbers confirm the low densities of the environments of our sample galaxies compared to group and cluster like environments.
\begin{table}
    \centering
    \caption{Local environment density estimates for the sample galaxies. Also included for comparison are the density estimates for a few clusters and groups. Here $d_{5}$ is the projected distance to the 5th-nearest neighbour and $\Sigma_{5}$ is the projected surface density corresponding to a radius of $d_{5}$.}
    \label{tab:densities}
    \begin{tabular}{lccc}
        \hline \hline
         Name & Type & $d_{5}$ & $\Sigma_{5}$ \\
             &  & [Mpc] & [Mpc$^{-2}$] \\
        \hline
        NGC 1350 & (R')SB(r)ab &  0.51 & 6.06 \\
        NGC 1543 & (R)SB0      &  0.30 & 17.1 \\
        NGC 1792 & SA(rs)bc    &  0.35 & 13.2  \\
        NGC 2369 & SB(s)a      &  0.85 & 2.18 \\
        NGC 6300 & SB(rs)b     &  0.81 & 2.45 \\
        UGCA 168 & Scd         &  0.39 & 10.65 \\
        \hline
        Fornax   & Cluster  & 0.06 & 490.02 \\
        Hydra    & Cluster  & 0.05 & 625.6   \\
        Pavo     & Group    & 0.09 & 209.85  \\
        Dorado   & Group    & 0.10 & 150.2    \\
        Telescopium & Group  &  0.18 & 48.9    \\
        Eridanus & Group  &  0.28 & 19.8  \\
        \hline
        \end{tabular}
\end{table}

\subsection{Observations}
\label{subsec:observations}
The observations were carried out with the Australia Telescope Compact Array (ATCA) between January and February 2017. Each source was observed for 12 hrs in both the 750C and 1.5D array configurations. This ensures both good sensitivity, as well as the spatial resolution necessary for kinematic modeling. Both the continuum and spectral line observations were done simultaneously using the Compact Array Broad-band Backend (CABB, Wilson et al.~\citeyear{wilson11}). The continuum band has a total bandwidth of 2048 MHz with a frequency resolution of 1 MHz, centered at 2.1 GHz.
The total bandwidth of the spectral line observations is 8.5 MHz with a channel width of $0.5$ kHz. The details of the observations are listed in Table.~\ref{tab:obs-details}. The observation strategy included 55 min on the source and 5 min on the phase calibrator each hour. The primary flux calibrator PKS 1934-638 was observed for 10 min at the start and end of the observation run.

\begin{table*}
    \centering
    \caption{Details of the observation. $\nu_0$ is the center frequency of the observation.}
    \label{tab:obs-details}
    \begin{tabular}{lcccc}
        \hline \hline
         Name & $\nu_{0}$ & Phase calibrator & Synthesized beam & rms noise  \\ 
              & [MHz]    & & [arcsec]  & [mJy beam$^{-1}$]   \\      
        \hline
          NGC 1350  & 1415.50 & PKS 0237--233  & $57 \times 22$ & 1.2  \\
          NGC 1543  & 1419.00 & PKS 0420--625  & $38 \times 30$ & 1.4  \\
          NGC 1792  & 1419.00 & PKS 0438--436  & $44 \times 24$ & 1.6 \\
          NGC 2369  & 1409.00 & PKS 0823--500, PKS 0637--752  & $31 \times 24$ & 3.9 \\
          NGC 6300  & 1419.50 & PKS 1718--649  & $33 \times 23$ & 1.2 \\
          UGCA 168  & 1420.50 & PKS 0834--196  & $53 \times 25$ & 3.3  \\ 
        \hline
    \end{tabular}
\end{table*}

\subsection{Data reduction}
\label{subsec:data-reduction}

The data has been reduced using the \texttt{MIRIAD} software package (Sault et al.~\citeyear{sault95}). First,  the radio frequency interference is removed via an automatic pipeline. Next, the bandpass, flux and phase calibration are performed using the tasks \texttt{MFCAL}, \texttt{GPCAL} and \texttt{MFBOOT}. Continuum subtraction to the spectral data is then performed using the task \texttt{UVLIN}. The task \texttt{INVERT} is then used to combine both the 750C and 1.5D array data to make a dirty cube. The task \texttt{CLEAN} is run to clean the cube and restored using the task \texttt{RESTOR}. The cubes were made with the Brigg's robust parameter set to 0.5, to get a good balance of both sensitivity and resolution. The typical synthesized beam sizes are between 20 and 50 arcsec. The channels were also smoothed to a width of 5 \kms , which allows for reliable kinematic fits to the data. Finally, moment 0 (intensity) and moment 1 (velocity) maps were made for all six galaxies. The typical 3$\sigma$ noise levels in each channel in the cube is $\sim$ 3 -- 10 mJy beam$^{-1}$. The 3$\sigma$ noise levels in the moment 0 maps is $\sim$ 30 mJy beam$^{-1}$ \kms which corresponds to column densities $\sim N_{\textrm{\h1}} \sim 4 \times 10^{19}$ cm$^{-2}$. The next stage of the analysis is to fit 3D tilted ring models to the data cubes to extract useful kinematic properties.

\subsection{3D fitting to data}
\label{subsec:3dfit}

The \h1 emission in the reduced data cube is fit with a 3D tilted ring model to derive important kinematic and geometrical parameters associated with the galaxies, such as their rotation curves V$_{rot}$, dispersion velocities, inclination and position angles and surface brightness profiles.
The tilted ring model was first introduced by Rogstad et al. (\citeyear{rogstad74}), where they fit a tilted ring model to the M83 galaxy. The method is based on the premise that gas, stellar and other baryonic material are moving about the galactic centre in thin circular rings or annuli, where the rings may have varying inclination and position angles. A number of models are then built with varying ring inclination and position angles and a model that best represents the observed data is chosen as the best fit model.
Previously, tilted ring models were fit to 2D velocity maps of galaxies to infer their kinematics, however, this method is prone to error due to beam smearing effects in galaxies that are more edge on. 
Direct 3D fitting to spectral data cubes is more robust.
We use 3DBarolo (Di Teodoro et al.~\citeyear{teodoro15}), a 3D kinematic fitting code, to make fits to the \h1 data cubes. The code is input with initial guesses for the parameters to be fit, such as the geometrical centres of the rings, inclination and position angles, rotation velocities, dispersion velocities and the systemic velocity of the galaxy. The code then simulates model data cubes using a Monte-Carlo distribution of point sources which are then projected onto data cubes and convolved with an appropriate instrumental function (the synthesized beam). The model and the observed data cubes are compared and the model that yields a minimized $\chi^{2}$ value is chosen as the best fit model.
The code outputs the best fit rotation curve, surface brightness profile as well as inclination and position angles of all the concentric rings. 
The values and properties of the fit rings are used to compute the \h1 mass, stellar mass, specific and total angular momentum of the galaxies, the details of which are described in the following section. The moment 0 and moment 1 maps of the data, model and the residual (data - model) for all six galaxies are included in Appendix~\ref{App2:galaxy-models}. Also included are the best fit rotation curves, inclination and position angles as well as the \h1 surface density profiles.


\section{Specific angular momentum and \h1 gas}
\label{sec:angular-mom}

In this section we discuss the connection between the neutral atomic mass fraction and specific angular momentum of disc galaxies. \citetalias{obreschkow16} find a tight relation between the atomic mass fraction $f_{atm} = \frac{1.35M_{\textrm{\h1}}}{M}$ and the stability parameter $q = \frac{j\sigma}{GM}$, where $M_{\textrm{\h1}}$ is the \h1 mass of the galaxy, $j$ is the specific baryonic angular momentum, $\sigma$ is the dispersion velocity of the Warm Neutral Medium (WNM), $M$ is the total baryonic mass (stars + interstellar medium) and $G$ is the universal gravitational constant.
The authors describe the stability parameter as a global analog to the local Toomre stability parameter of a hypothetical single-component WNM disc. 

We now test this model using our 6 \h1-deficient galaxies. In order to get a good estimate on the angular momentum, we need trustworthy kinematic fits. Once a good 3D fit has been established, the 3D tilted rings are projected onto the 2D intensity maps of the galaxies. To calculate the \h1 mass we use the moment 0 (integrated flux) maps and to calculate the stellar mass, we use 2MASS (Skrutskie et al.~\citeyear{2mass06}) $K_{s}$-band background subtracted images of the galaxies. The $K_{s}$-band magnitudes are converted to stellar masses using the relation in Wen et al. (\citeyear{wen13}) (with $\log M/L \sim 1.105$ [M$_{\odot}$/L$_{\odot}])$. The \h1 and stellar mass is calculated within each tilted ring and summed up across all rings to get the total mass. Since we have the rotation curve for the galaxy from the 3D fit, we use the corresponding velocities at each ring and compute the angular momentum in each ring. The specific angular momentum is then calculated by dividing the total angular momentum by the total baryonic mass of the galaxy. 

We follow the definition prescribed by \citetalias{obreschkow16} and define the atomic fraction as:
$f_{\small atm} = 1.35 M_{\small \textrm{\h1}}/M$, where $M = M_{\star} + 1.35(M_{\small \mathrm{\textrm{\h1}}} + M_{\small \mathrm{H_{2}}})$ is the total baryonic mass and  $M_{\mathrm{\small \textrm{\h1}}}$, $M_{\star}$ and $M_{\small \mathrm{H_{2}}}$ are the \h1, stellar and molecular hydrogen mass respectively. The factor 1.35 accounts for the 26\% He in the local universe. $M_{\small \mathrm{H_{2}}}$ is not directly available for the sample galaxies and so we have assumed a reasonable mass ratio of $M_{\small \mathrm{H_{2}}}/M \sim 4\%$ following Obreschkow \& Rawlings (\citeyear{obreschkow09}). As mentioned earlier the atomic stability parameter is given by $q = \frac{j \sigma}{GM}$, where $j=\frac{\sum_{i} (1.35M_{\mathrm{\tiny \textrm{\h1}},i}+M_{\star,i})V_{rot,i}r_{i}}{\sum_{i} (1.35M_{\mathrm{\tiny \textrm{\h1}},i}+M_{\star,i})}$, here $r_{i}$ is the radius corresponding to the $i^{th}$ ring and $V_{rot,i}$ is the rotation velocity corresponding to that ring. 
The \h1 dispersion velocities ($\sigma$) of the galaxies have been measured from the 3D fitting to data. The median values of the same are given in Table.~\ref{tab:results}.
The results from the angular momentum analysis for the \h1-deficient galaxies are presented and discussed in Section~\ref{subsec:fHI-q}


\section{Results and Discussion}
\label{sec:results}

\subsection{General properties of the galaxies}
\label{subsec:general-properties}

\begin{figure*}
\hspace*{-1.2cm}
\includegraphics[width=18cm,height=12cm]{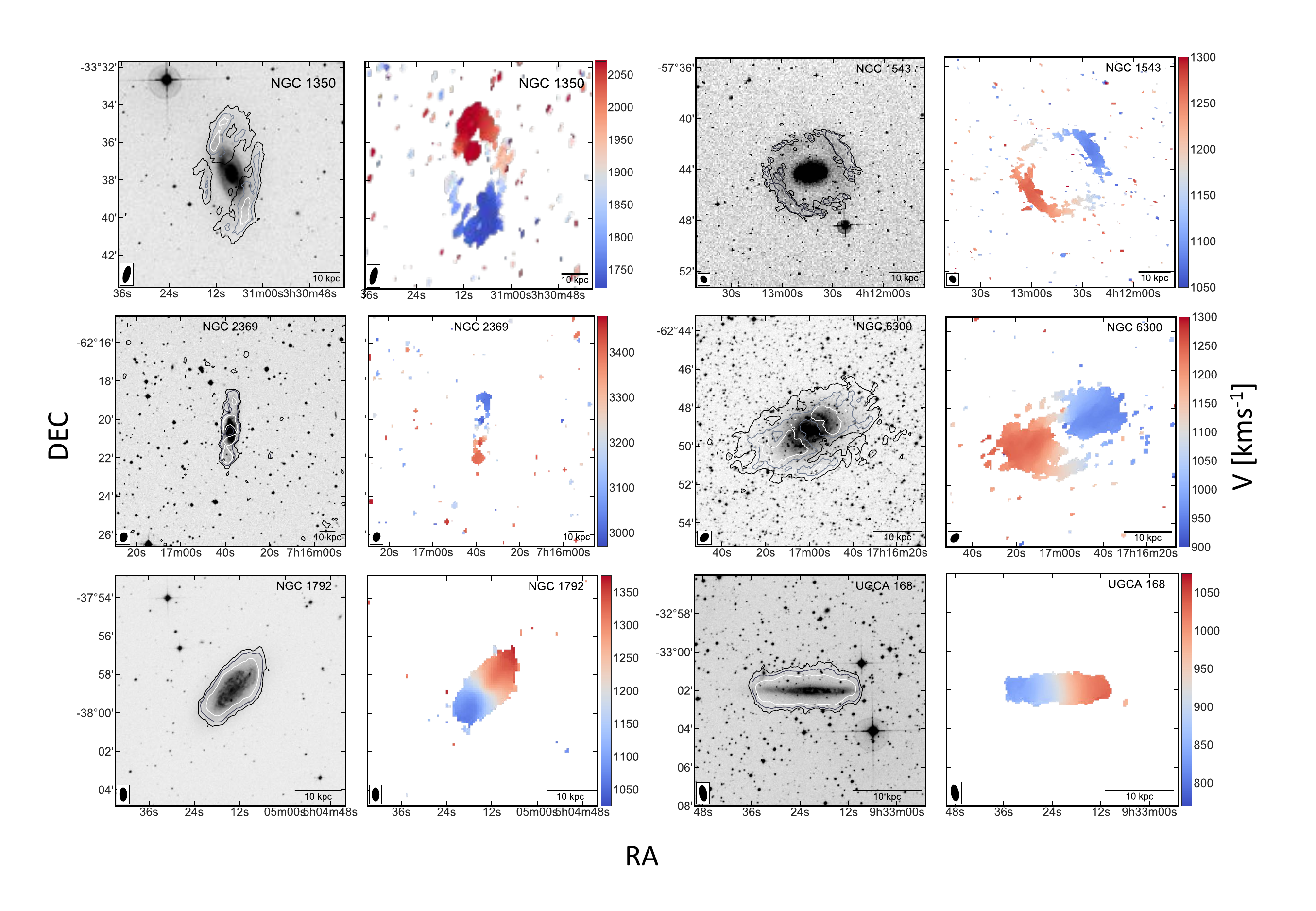}
\caption{The sample: The \h1 contours (black, grey and white are set at $3 \times 10^{20} cm^{-2}$, $5 \times 10^{20} cm^{-2}$ and $1 \times 10^{21} cm^{-2}$ for NGC 1792, NGC 2369 and UGCA 168. For NGC 1350 and NGC 6300 the contours are set to $1 \times 10^{20} cm^{-2}$, $3 \times 10^{20} cm^{-2}$ and $5 \times 10^{20} cm^{-2}$ and lastly for NGC 1543 the contours are set at $5 \times 10^{19} cm^{-2}$, $1 \times 10^{20} cm^{-2}$ and $3 \times 10^{20} cm^{-2}$ ) are overlaid on top of the DSS optical images. Also shown are their corresponding velocity maps. Clearly seen for many galaxies in the sample are the predicted central \h1 holes for their low values of $q$.}
\label{fig:galaxy-maps}
\end{figure*}

In this section we discuss the general properties of the sample galaxies. Fig.~\ref{fig:galaxy-maps} shows the \h1 maps of the galaxies overlaid on top of their DSS optical images. We notice that their \h1 discs extend only so far as their stellar discs, which is typical of \h1-deficient spirals (see for example D\'{e}nes et al.~\citeyear{denes16}). The galaxies in our sample have optical morphologies that range from Sb0 to Sc. NGC 1350, NGC 1543 and NGC 6300 are more bulge like in their centers and also appear to possess bars. In terms of their \h1 morphology, we observe \h1 hole like features in the central regions of their discs. The \h1 holes correspond to regions dominated by stellar and molecular material. One of the predictions of the theoretical models in \citetalias{obreschkow16} is the presence of these \h1 holes, whose sizes also correlate with their $q$ values. We discuss the holes in more detail in Section~\ref{subsection:HIholes}. 

The \h1 morphology of the galaxies do not show any asymmetric/lopsided or disturbed features, which are commonly observed among galaxies affected by ram pressure. Other typical signatures of ram pressure include truncated discs and enhanced star formation in the leading side of the galaxy (Cayatte et al.~\citeyear{cayatte94}; Vollmer et al.~\citeyear{vollmer01}). Since we do not observe such features in our sample, we believe ram pressure may only be playing a minimal role in affecting their \h1 content, if any. Moreover, none of the galaxies in our sample reside in cluster-like environments for ram pressure to take effect. We also do not see any strong evidence for tidal tails, streams and/or warps in the outer regions of the discs of the sample galaxies, that may allude to previous tidal interactions (except NGC 1543, whose case is discussed in more detail below).

A peculiar galaxy in the sample is NGC 1543, where we observe a ring of \h1 gas surrounding the central nucleus (see Fig.~\ref{fig:galaxy-maps}). The ring is also observed in the optical as well as the UV, indicating current star formation. The origin of such a symmetric ring is unclear, however, NGC 1543 is at the very edges of the Dorado group and as such, it is possible that it is a ``backsplash" galaxy, meaning it has passed through the group at least once and along the way collided with a dwarf galaxy, driving the formation of the ring. Simulations support this scenario and indicate that an interaction via collision or drop-through is the most common way such ring galaxies are formed (see for example Elagali et al.~\citeyear{elagali18}). NGC 1543 is classified as an S0 galaxy and does not seem to possess a normal disc. However, we have used the rotation information from the 3D fitting to the \h1 data and computed the $q$ value for this galaxy. The individual galaxies and their environments are discussed in more detail in Appendix~\ref{App:Env}.

\subsection{The $f_{\small atm} - q$ relation}
\label{subsec:fHI-q}

\begin{figure}
\hspace*{-0.6cm}
\includegraphics[width=8.2cm,height=6.2cm]{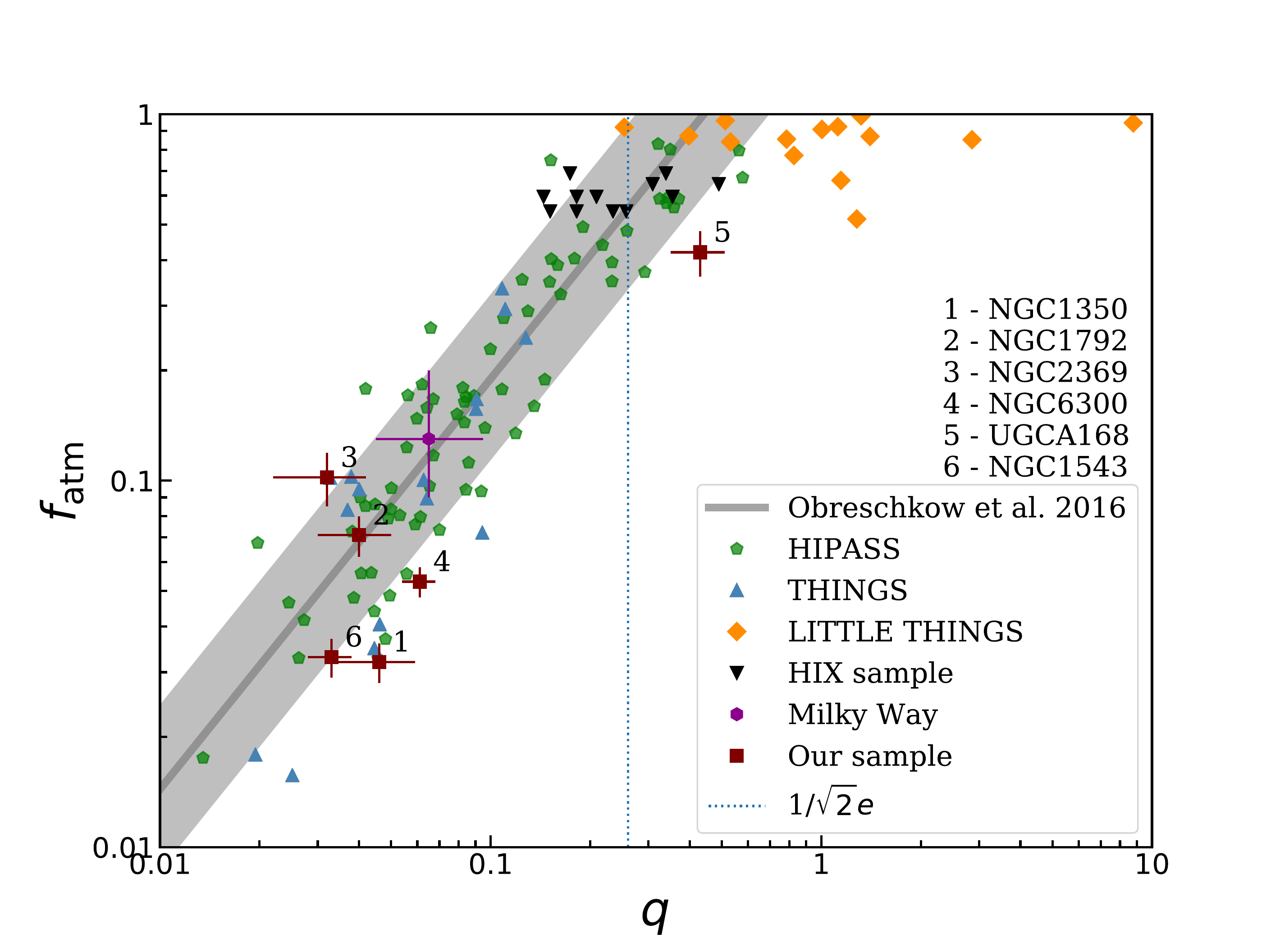}
\caption{The $f_{atm} - q $ relation. The galaxies in our sample are represented by the maroon squares. The dark gray line is the analytical model for $f_{atm}$ from \citetalias{obreschkow16}. The shaded gray region shows the 40\% scatter about the model. Also shown are galaxies from THINGS, LITTLE THINGS, HIPASS and the \h1X surveys. The vertical dotted line represents $ q = 1/\sqrt{2}e$, the threshold beyond which axially symmetric exponential disks of constant velocity can remain entirely atomic.}
\label{fig:fatmVsq}
\end{figure}

Fig.~\ref{fig:fatmVsq} shows the scaling relation between the atomic mass fraction $f_{\small atm}$ and the stability parameter $q$ for a sample of local disc-like galaxies. The plot shows galaxies from THINGS (Walter et al.~\citeyear{things08}) for which the $f_{\small atm}$ and $q$ values have been calculated by Obreschkow \& Glazebrook (\citeyear{obreschkow14})[hereafter \citetalias{obreschkow14}] and LITTLE THINGS survey (Hunter et al.~\citeyear{littlethings12}) for which $j$ values were measured by Butler et al. (\citeyear{butler17}) and $q$ values calculated by \citetalias{obreschkow16}. Also shown are galaxies from the HIPASS survey (Meyer et al.~\citeyear{hipass04}), for which $f_{\small atm}$ and $q$ values have been estimated by \citetalias{obreschkow16} as well as the \h1-excess galaxies from the \h1X survey (Lutz et al.~\citeyear{lutz17},~\citeyear{lutz18}). Also included is the Milky Way, for which the $f_{atm}$ and $q$ values have been calculated by \citetalias{obreschkow16}. We find that the galaxies in our sample follow the relation fairly consistently but with some scatter. All six galaxies in our sample (within errors) fall within the 40\% scatter (light shaded region) about the analytic model (thick grey line in Fig.~\ref{fig:fatmVsq}) from \citetalias{obreschkow16}. 
This result comes in support of the idea that angular momentum plays an important role in regulating the \h1 gas content in disc galaxies, particularly those from low-density environments. Galaxies with lower specific angular momentum have lower $q$ values which directly sets instabilities in the disc leading to increased star formation. This tends to deplete the \h1 gas reservoir within about 2 -- 3 Gyrs (Schiminovich et al.~\citeyear{schiminovich10}; Leroy et al.~\citeyear{leroy08}; Genzel et al.~\citeyear{genzel10}). On the contrary, galaxies with higher specific angular momenta will support a more stable disc leading to a reduced star formation efficiency. These systems will appear to have excess \h1 for their given \textit{R}-band magnitudes (or stellar mass).
 
It is worth noting that UGCA 168 does not appear to be \h1 deficient as can be seen by its location in the $f_{\small atm} - q$ relation.
There may be a number of reasons for this. Firstly, this discrepancy may be associated with any uncertainty in the measured \textit{R}-band magnitude for this galaxy. We have used the \textit{R}-band magnitudes for our sample galaxies derived by D\'{e}nes et al. (\citeyear{denes14}) from HOPCAT (Doyle et al.~\citeyear{doyle05}), which makes use of the SuperCOMOS data (Hambly et al.~\citeyear{supercosmos01}). The standard error in the \textit{R}-band for SuperCOSMOS is close to 0.6 mag. This uncertainty could potentially shift the position of this galaxy in the $M_{\textrm{\h1}} ~-$ M$_{R}$ relation (Fig.~\ref{fig:sample}) and categorize it is as \h1-normal and not \h1-deficient. Additionally, it is possible that the internal dust extinction corrections made by D\'{e}nes et al. may not have properly accounted for the reddening of the M$_{R}$ magnitude, displacing the galaxy in the $M_{\textrm{\h1}} ~-$ M$_{R}$ scaling relation and leading to its erroneous classification as \h1-deficient. Secondly, UGCA 168 is highly inclined and almost edge-on. This makes it difficult even for 3D fitting codes to reliably fit tilted-ring models to the data. This may lead to inconsistencies in the measurements, particularly the dispersion velocity, which is affected by beam smearing effects and is sensitive to the inclination of the galaxy. The somewhat higher dispersion velocity ($\sim 16$~\kms) we measure for this galaxy is likely a residual beam smearing effect. 

It is also interesting to note that the ring galaxy NGC 1543 falls on the relation. However, we point out that the calculated $q$ value for this galaxy may not be accurate. This is because the measured rotation velocities from the 3D fitting is unreliable in the inner regions, where the S/N of the \h1 emission is low. Additionally, the galaxy supports a ring and not a normal disc. This makes it more difficult to understand the behaviour of this galaxy on the $f_{\small atm} - q$ plane.
We would need a larger sample of ring galaxies to see if they consistently follow the relation. This may possibly be explored in a future work.

Table.~\ref{tab:results} lists the fit parameters, as well as the atomic mass fraction ($f_{\textrm{atm}}$), $j$ and $q$ values for our sample.

\begin{table*}
    \centering
    \caption{Details of the fit parameters from 3DBarolo along with the neutral atomic mass fraction and $q$ values for the sample. $V_{\textrm{flat}}$ is the maximum constant rotation velocity. Median values for $\sigma_{\textrm{\h1}}$, $i$ and PA have been quoted. $j$ is the total baryonic specific angular momentum.}
    \label{tab:results}
    \begin{tabular}{lccccccc}
        \hline \hline
         Name & $V_{\textrm{flat}}$ & $\sigma_{\textrm{\h1}}$ & $i$ & PA & $f_{atm}$ & $\log(j)$ & $q$ \\
             & [\kms] & [\kms] & [deg] & [deg] &  & [M$_{\odot}$ \kms kpc] &  \\
        \hline
        NGC 1350 & 217 & 10.5  & 59 & 15   & 0.032  & 3.18  & 0.046 \\
        NGC 1543 & 180 & 5.9   & 24 & 134  & 0.033  & 2.77  & 0.033 \\
        NGC 1792 & 147 & 10.1  & 62 & 318  & 0.071  & 2.70  & 0.04  \\
        NGC 2369 & 200 & 12.2  & 77 & 169  & 0.11   & 2.96  & 0.032 \\
        NGC 6300 & 178 & 12.1  & 51 & 111  & 0.053  & 2.83  & 0.061 \\
        UGCA 168 & 100 & 16.2  & 79 & 271 & 0.42   & 2.55  & 0.43  \\
         \hline
     \end{tabular}
\end{table*}

\subsection{\h1 holes and $q$ values}
\label{subsection:HIholes}

\begin{figure}
\hspace*{-0.6cm}
\includegraphics[width=8.2cm,height=6.2cm]{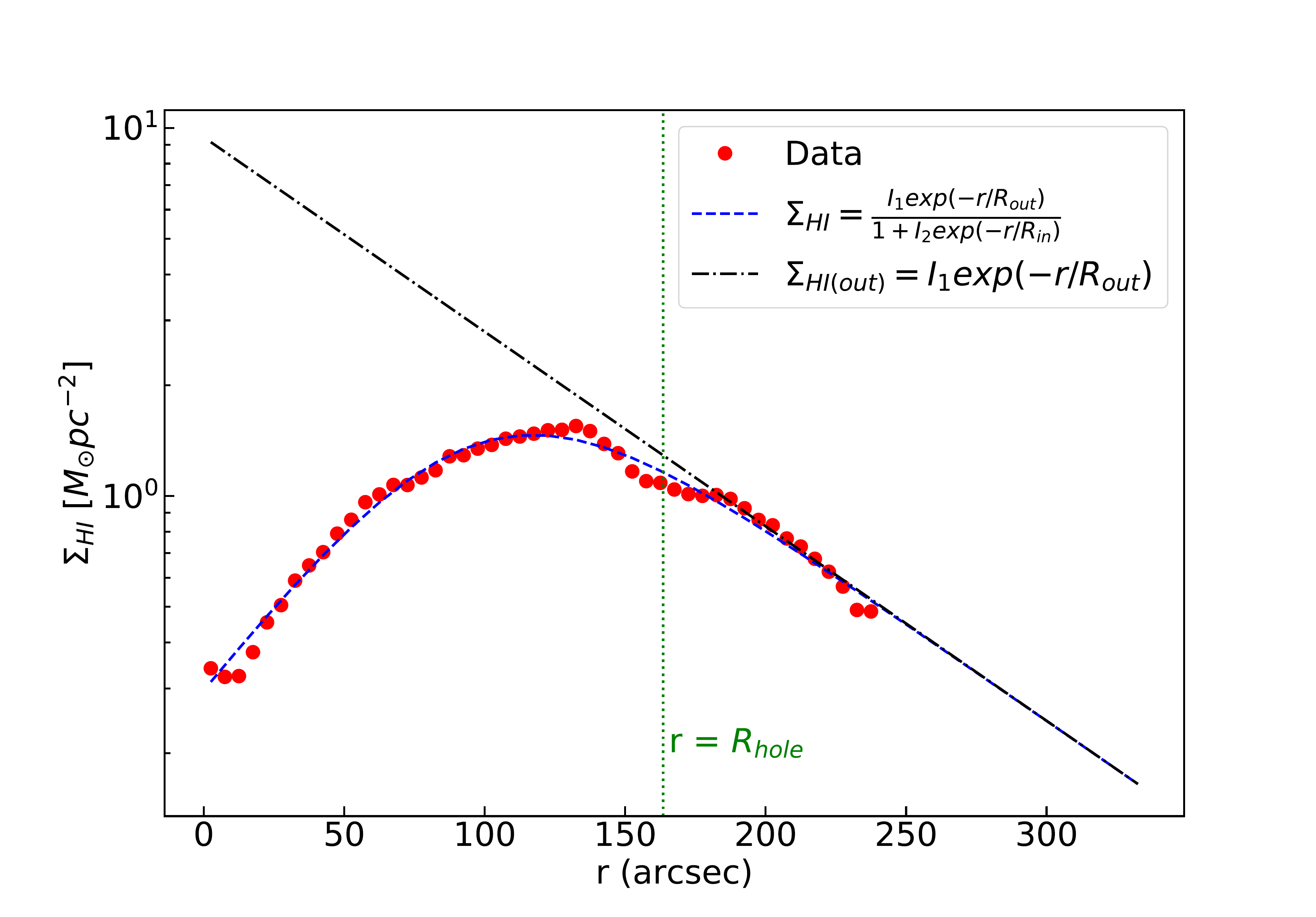}
\caption{The plot shows shows the method employed to measure the \h1 hole size $R_{hole}$ for the galaxy NGC 1350. The blue dashed line is a two component exponential fit to the \h1 surface density profile ($\Sigma_{\textrm{\h1}}$) of the galaxy. The dot-dash line is the exponential fit to $\Sigma_{\textrm{\h1}}$ corresponding to the outer parts of the disc. The vertical dotted green line represents the radius at which $\Sigma_{\textrm{\h1}}$ deviates by 10\% from the $\Sigma_{\textrm{\h1}(outer)}$.}
\label{fig:Rhole-example-fit}
\end{figure}

Since we are looking at \h1-deficient galaxies from low-density environments with no close neighbours to affect their \h1 distribution, we are in a position to test a key prediction made by \citetalias{obreschkow16}.
The analytic models predict the existence of \h1~``holes" in the discs of galaxies for which $q < 1/\sqrt{2}e$. They describe the radius at which the disc transitions from the stellar and/or molecular gas dominated part to the \h1 dominated part as the size of the \h1 hole ($R_{hole}$). Such \h1 holes in disc galaxies have long been observed (see for example Wong \& Blitz~\citeyear{Wong02}). Blitz \& Rosolowsky (\citeyear{blitz04}) suggest that the mid-plane hydrostatic pressure in the discs of galaxies regulate the distribution of \h1 and \hh ~and therefore naturally define a transition radius. Bigiel \& Blitz (\citeyear{bigiel12}) also find results in support of this argument. However, the \citetalias{obreschkow16} study link the size of the \h1 holes to the specific angular momentum of galaxies for the first time.
Given the simple assumptions made in their model, a $q$ value greater than $1/\sqrt{2}e$ enforces complete disc stability (with the local Toomre parameter for atomic gas Q$_{atm} > 1$ for all regions in the disc), thereby making the entire disc fully atomic. However, if the $q$ value for a given disc galaxy is less than this threshold, the model predicts that the inner/core regions will be filled with only molecular and stellar material, thereby creating \h1 depressions or ``holes". In the idealised model of \citetalias{obreschkow16}, the \h1 goes to zero in the central parts, the discs of real galaxies however will transition from \h1 dominated outer parts to non-atomic dominated central parts at a break radius. We attempt to measure this radius by profile fitting the \h1 surface density and see if it correlates with the prediction of the $f_{\small atm} - q$ model.
The relation for the \h1 hole size is given by $R_{hole} = X_{hole} \cdot R_{disc}$, where $X_{hole}$ is derived from the analytic model that describes the neutral atomic mass fraction ($f_{atm}$) and $q$, and $R_{disc}$ is the scale length (for details see \citetalias{obreschkow16}). We calculate $X_{hole}$ for our sample galaxies and the 16 THINGS galaxies from the \citetalias{obreschkow16} paper, as well as the Milky Way. $R_{disc}$ has been calculated by making an exponential fit to the disc stellar surface density profile, while omitting the central regions in the fit to avoid bulge and other non-disc components. To estimate $R_{hole}$, we first fit a two component exponential to the \h1 surface density profile (accounting for the central and outer parts of the disc) following Wang et al. (\citeyear{wang14}), and defined as follows:

\begin{equation}
    \Sigma_{\textrm{\h1}} = \frac{I_{1}\exp(-r/R_{out})}{1 + I_{2}\exp(-r/R_{in})}
\end{equation}

Where $I_{1}$, $R_{out}$, $I_{2}$ and $R_{in}$ are free parameters. At large radii the denominator reduces to one and only the component in the numerator remains, describing the outer $\Sigma_{\textrm{\h1}}$. We define the \h1 hole size as the radius at which the \h1 surface density deviates by 10\% from the outer \h1 exponential surface density. To illustrate this, an example is shown in Fig.~\ref{fig:Rhole-example-fit}. For the Milky Way, we have used the \h1 surface density distribution from Kalberla \& Dedes (\citeyear{kalberla08}) and followed the method described above to calculate $R_{hole}$ while assuming a scale radius, $R_{disc} \sim 3 \pm 1$ kpc (Kent et al.~\citeyear{kent91}; McMillan~\citeyear{mcmillan11}). 

As mentioned earlier, NGC 1543 has a \h1 ring which is likely to have formed due to a previous interaction as discussed in Section~\ref{subsec:general-properties}. Applying the above method to determine the \h1 hole size for this galaxy will wrongly identify the region inside the ring as the hole. The formation mechanism of the ring (due to an external agency) and the \h1 hole (due to intrinsic disc instabilities regulated by angular momentum) are entirely different and for this reason we have excluded NGC 1543 from our \h1 hole analysis.

\begin{figure}
\hspace*{-0.42cm}
\includegraphics[width=8cm,height=5.9cm]{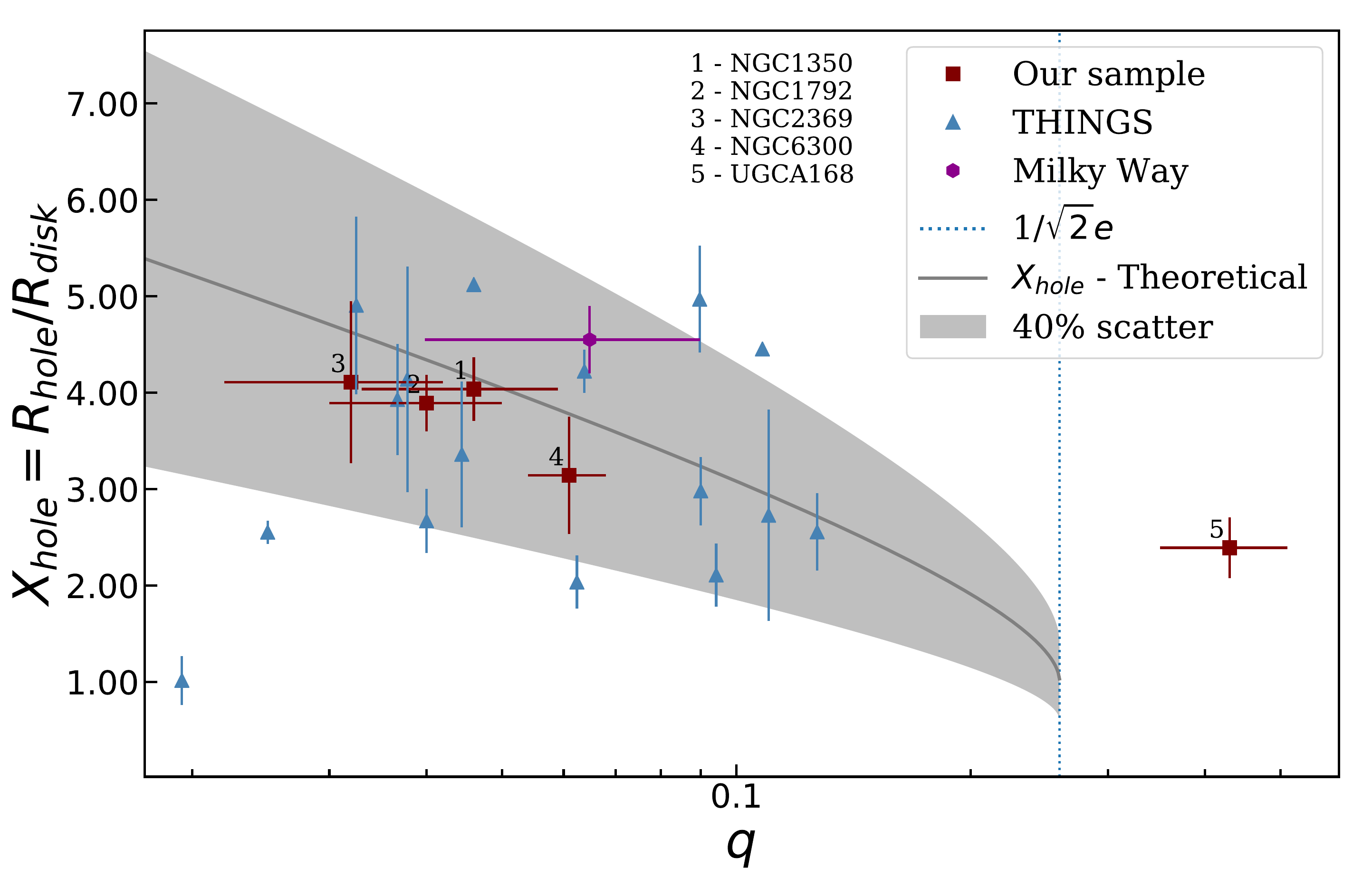}
\caption{Measured $X_{hole}$ values plotted against the $q$ parameter for our sample galaxies (maroon squares), the 16 THINGS galaxies (blue triangles) from \citetalias{obreschkow16} and the Milky Way. The gray solid line is the analytic relation of $X_{hole}$ as a function of $q$, described in \citetalias{obreschkow16}, with a 40\% scatter shown by the light gray shaded area. The dotted vertical line at $q = 1/\sqrt{2}e$ defines the threshold beyond which axially symmetric exponential disks of constant velocity can remain entirely atomic, remaining \h1 hole free.}
\label{fig:HIholeVsq}
\end{figure}

Fig.~\ref{fig:HIholeVsq} shows the measured $X_{hole}$ values plotted against the $q$ parameter. The galaxies follow the general trend predicted by the analytic model but with a large scatter. 
Most galaxies fall within the 40\% scatter about the analytic model. Reasons for the scatter maybe two-fold - one associated with effects that are intrinsic to the discs of galaxies and the other linked to the uncertainties arising from our very definition of \h1 holes and their associated measurement errors. In relation to the former, the \h1 hole/depression in the centers of spiral galaxies may not only be dictated by their specific angular momentum, but a number of other factors. Energetic processes in the centers of galaxies such as AGNs have the ability to ionize and blow out large fractions of gas material. Other features such as bars interact with gas at their edges which results in the loss of angular momenta of the gas material, due to which they begin to move towards the central regions. Such instabilities induced by the bars will result in increased star formation and hence the depletion of gas in the centres of galaxies, creating \h1 holes for an entirely different reason. Uncertainties may also be associated with the method we follow to determine the \h1 hole radius. As mentioned earlier, the \citetalias{obreschkow16} paper describes the \h1 hole radius as the transition radius, where \h1 begins to dominate over stellar and molecular gas. Therefore, in order to properly estimate $R_{hole}$, one needs to 
carefully model the \h1-to-non-atomic material ratios in the discs of galaxies to identify more clearly the \h1 dominated regions and the regions dominated by stars and molecular gas. Since our method involves only the use of the \h1 surface density to identify the \h1 holes, some caution is warranted when determining the \h1 hole size. For many stable disc galaxies, the peak \h1 surface density will roughly match the transition radius from the stellar dominated to \h1 dominated regions (such as NGC 1350 shown in Fig.~\ref{fig:Rhole-example-fit}), while in other cases, this may not be true. We leave a more detailed analysis of the same for a future work. Regardless of these other effects and caveats, overall, the galaxies follow the model prediction. The fact that such holes exist and correlate with the galaxies' $q$ values only strengthens our argument in favour of specific angular momentum playing an important role in regulating star formation in the centers of galactic discs, and the fraction of atomic gas they are able to retain. This also validates the theoretical predictions of \citetalias{obreschkow16}. Table.~\ref{tab:Xhole-q-all} lists the measured \h1 hole sizes for our sample, a few THINGS galaxies and the Milky Way.

\begin{table}
    \centering
    \caption{\h1 hole sizes for our sample, a few THINGS galaxies as well as the Milky Way. Here $R_{disc}$ is the exponential disc scale length and $R_{hole}$ is the \h1 hole size measured using the method described in Section~\ref{subsection:HIholes}. $X_{hole}^{theory}$ have been estimated using the analytic models in \citetalias{obreschkow16}.}
    \label{tab:Xhole-q-all}
    \begin{tabular}{lccccc}
        \hline \hline
         Name & $R_{disc}$ & $R_{hole}$ & $X_{hole}^{obs}$ & $X_{hole}^{theory}$ & $q$ \\
              & [kpc]      & [kpc]      &                  &         & \\ 
        \hline
        NGC1350	&	4.81	&	19.43	&	4.04	&	4.53	&	0.046	\\
        NGC1792	&	2.14	&	8.33	&	3.89	&	3.83	&	0.04	\\
        NGC2369	&	3.03	&	12.46	&	4.11	&	5.20	&	0.032	\\
        NGC6300	&	2.27	&	7.14	&	3.14	&	4.05	&	0.061	\\
        UGCA168	&	2.02	&	4.84	&	2.39	&	1.34	&	0.43	\\
        \hline
        THINGS &     &           &           &           &           \\
        \\
        NGC628  &   2.3     &   6.86    &   2.39    &   1.34    &   0.27 \\
        NGC925	&	3.46	&	9.44	&	2.73	&	2.94	&	0.11 \\
        NGC2403	&	1.10	&	2.82	&	2.56	&	2.69	&	0.13 \\
        NGC2841	&	4.77	&	9.71	&	2.04	&	3.73	&	0.06 \\
        NGC2976	&	1.06	&	2.24	&	2.11	&	3.15	&	0.09 \\
        NGC3184	&	2.38	&	10.06	&	4.22	&	3.73	&	0.06 \\
        NGC3198 &   3.2     &   14.25   &   4.45    &   2.94    &   0.11 \\
        NGC3351 &   2.2     &   7.38    &   3.36    &   4.18    &   0.04 \\
        NGC3521	&	3.80	&	18.63	&	4.91	&	4.59	&	0.03 \\
        NGC3627 &   2.8     &   7.15    &   2.55    &   4.92    &   0.025 \\  
        NGC4736	&	1.06	&	1.07	&	1.01	&	5.24	&	0.02 \\
        NGC5055	&	3.18	&	13.16	&	4.14	&	4.18	&	0.04 \\
        NGC5194 &   2.8     &   14.34   &   5.12    &   4.14    &   0.046 \\
        NGC6946 &   2.5     &   6.67    &   2.67    &   4.18    &   0.04 \\
        NGC7331	&	4.19	&	16.47	&	3.93	&	4.43	&	0.037 \\
        NGC7793 &   1.3     &   6.46    &   4.97    &   3.15    &   0.09 \\
         \hline
        Milky Way & 3       &   13.65   &   4.55    &   3.67    &   0.065   \\
        \hline
     \end{tabular}
\end{table}

\section{Conclusions}
\label{sec:Discussion}

In this study we have computed the specific baryonic angular momentum of six late type galaxies and examined its influence on their \h1 gas mass fractions. The galaxies were selected on the basis that they are \h1-deficient in the $M_{\textrm{\h1}} ~-$ M$_{R}$ relation. They have also been sampled from low-density environments in order to minimize the effects of ram pressure stripping and tidal interactions. 
We have used 3DBarolo, a 3D tilted ring fitting code, to derive the kinematic properties of the galaxies including their inclination and position angles, rotation and dispersion velocities and surface brightness profiles. We find that the galaxies in our sample follow the $f_{atm} - q$ relation as predicted by Obreschkow et al. (\citeyear{obreschkow16}), without systematic offset. 
While historically, \h1 deficiency in galaxies has often been associated with the influences of the environment in removing gas, this result brings to light the role of specific angular momentum in regulating the \h1 gas among galaxies in low-density environments. 
Angular momentum by virtue of its cosmic variance, will play an important role in controlling the evolution of galaxies, with galaxies having higher specific baryonic AM retaining a larger fraction of their gas and appearing \h1-excess and those with lower specific AM depleting gas more efficiently and appearing gas-depleted. This effect will naturally contribute to the scatter that is observed in the various scaling relations. However, in order to verify this, we need accurate values of $q$ measured for a large number of galaxies. By identifying the effect of angular momentum on scaling relations and accounting for it, we may be able to reduce or constrain the observed scatter. This calls for the need to have well resolved \h1 maps for potentially thousands of galaxies. With upcoming large \h1 surveys such as the WALLABY survey (Koribalski~\citeyear{koribalski12}) using ASKAP, APERTIF survey (Oosterloo et al.~\citeyear{oosterloo9}) and eventually the SKA, this may soon become a reality.\\
\indent A consequence of the analytic models developed in \citetalias{obreschkow16} to derive the $f_{atm} - q$ relation is the presence of central \h1 ``holes" in the discs of galaxies. The \h1 moment maps of our sample galaxies (Fig.~\ref{fig:galaxy-maps}) show clear indications of the presence of such holes. We have introduced a heuristic method to estimate the sizes of the \h1 holes, and applied the method to measure the hole sizes in our sample galaxies, few galaxies from THINGS as well as the Milky Way. We find that the measured \h1 hole sizes correlate well with the model prediction to within 40\% scatter. This complementary result comes in support of the theoretical prediction that specific angular momentum plays a dominant role in regulating the \h1 gas content in disc galaxies.

\section*{Acknowledgements}
\label{sec:acknowledgements}
We would like to thank the referee for useful comments and suggestions.
CM is supported by the Swinburne University Postgraduate Award (SUPRA). CM would further like to thank Chris Fluke, Deanne Fisher, Alister Graham, Tiantian Yuan and Michelle Cluver for their useful comments and discussions. \\
This publication makes use of data products from the Two Micron
All Sky Survey, which is a joint project of the University
of Massachusetts and the Infrared Processing and Analysis Center/
California Institute of Technology, funded by the National Aeronautics
and Space Administration and the National Science Foundation. \\
The Australia Telescope Compact Array is part of the Australia
Telescope National Facility which is funded by the Australian Government
for operation as a National Facility managed by CSIRO. \\
This research has made use of the NASA/IPAC Extragalactic
Database (NED), which is operated by the Jet Propulsion Laboratory,
California Institute of Technology, under contract with the
National Aeronautics and Space Administration. \\
This research has made use of \texttt{python} \url{https://www.python.org} and python packages: \texttt{astropy} (Astropy Collaboration et al.~\citeyear{astropy13}), \texttt{matplotlib} \url{http://matplotlib.org/} (Hunter et al.~\citeyear{Hunter:2007}), \texttt{APLpy} \url{https://aplpy.github.io/}, \texttt{NumPy} \url{http://www.numpy.org/} (van der Walt et al.~\citeyear{walt11}) and \texttt{SciPy} \url{https://www.scipy.org/} (Jones et al.~\citeyear{scipy01}). 




\bibliographystyle{mnras}
\bibliography{ref}


\appendix
\section{Environments and Morphology}
\label{App:Env}

We discuss and comment on the environments and morphologies of the individual galaxies in our sample. The quoted optical morphologies are based on the classification made by de Vaucouleurs et al. (\citeyear{devaucouleurs91}). The galaxies in our sample span the extent of the late type morphologies along the Hubble fork, from Sa to Sc. This is helpful, as it allows us to test if angular momentum and/or \h1 mass fractions show variations with their observed morphologies.\\

\textit{NGC 1350} is classified as (R')SB(r)ab and lies in the outskirts of the Fornax Cluster, at a distance of $\sim$ 7 Mpc ($\sim$ 5R$_{200}$) from the cluster centre. This rules out the possibility of ram pressure removing the \h1 gas from the galaxy. As per the nearest neighbour estimates, we do not find any close neighbours to this galaxy that may have interacted with it in the recent few Gyrs. We also do not find any evidence of tidal interactions, which if any, should have shown in its \h1 disk profile. This galaxy has very little \h1 in its central parts, almost as if there is an \h1 ``hole" (see Fig.~\ref{fig:galaxy-maps}). We discuss the existence of such \h1 holes in the context of a galaxy's specific angular momentum in more detail in Section~\ref{subsection:HIholes}.\\ 

\textit{NGC 1543} is classified as (R)SB0. A prominent ring is visible in both the optical and the \h1 about the nucleus of the galaxy. NGC 1543 resides in the outskirts of the Dorado group (at a projected distance $\sim$ 410 kpc or 0.7R$_{200}$). Dorado, is a loose group with over 26 confirmed members (Kilborn et al.~\citeyear{kilborn05}). Osmond and Ponman (\citeyear{osmond04}) made X ray observations of the Dorado group and find no group scale emissions, indicating that the group is relatively young and has not virialised yet. This again rules out ram pressure as a gas removing agent.
No evidence of diffuse extended \h1 is found, indicating that the galaxy does not have any ongoing interactions. \\

\textit{NGC 1792} is part of the ``NGC 1792 group" or the ``NGC 1808 group" - a loose group. 
It is about 160 kpc ($\sim$ 0.5R$_{200}$) from the group centre. This galaxy falls under the morphological classification SA(rs)bc. 
Dahlem et al. (\citeyear{dahlem01A}) looked for diffuse extended intergalactic \h1 gas in the form of tidal debris and tails between NGC 1792 and NGC 1808 but find no evidence of such features. Our \h1 maps also do not show signs of a disturbed \h1 disc. Since we do not find any clear evidence of ram pressure or tidally induced gas stripping, we conclude that such environmental parameters may not be playing a significant role in affecting the \h1 gas content of this galaxy. \\

\textit{NGC 2369} is a Luminous Infrared Galaxy (LIG) and classified as a SB(s)a. The optical image of the galaxy shows the presence of a bar and two prominent arms. It is an isolated galaxy with the nearest group (LDCE 0505) at a projected distance of $\sim$ 0.7 Mpc. We find no diffuse and extended \h1 gas surrounding this galaxy thereby ruling out any recent tidal interactions, and the low densities of the IGM will not support ram pressure stripping. \\

\textit{NGC 6300} is classified as SB(rs)b and is known to be a Seyfert type 2 galaxy. Ryder et al. (\citeyear{ryder96}) studied the \h1 distribution in NGC 6300 using the ATCA and find an extended tail like feature in the South-west corner of the galaxy. We also notice this feature as shown in Fig.~\ref{fig:galaxy-maps}. However, since there are no nearby neighbours that are likely to have caused this tidal tail like feature, Ryder et al. conclude that this could be a sign that NGC 6300 is still accreting \h1 gas from the IGM. \\

\textit{UGCA 168} is an Scd type galaxy residing in the outskirts ($\sim$ 470 kpc or 2R$_{200}$) of the group LGG 180. This is a loose group with 12 members (Pisano et al.~\citeyear{pisano11}). Pisano et al. note that they do not observe any intra-group X-ray emission in the LGG 180 group, thus indicating that the IGM is not dense enough for ram pressure to take effect. 

\section{Galaxy models}
\label{App2:galaxy-models}
\begin{figure*}
\hspace*{-1.4cm}
\includegraphics[width=20cm,height=16cm]{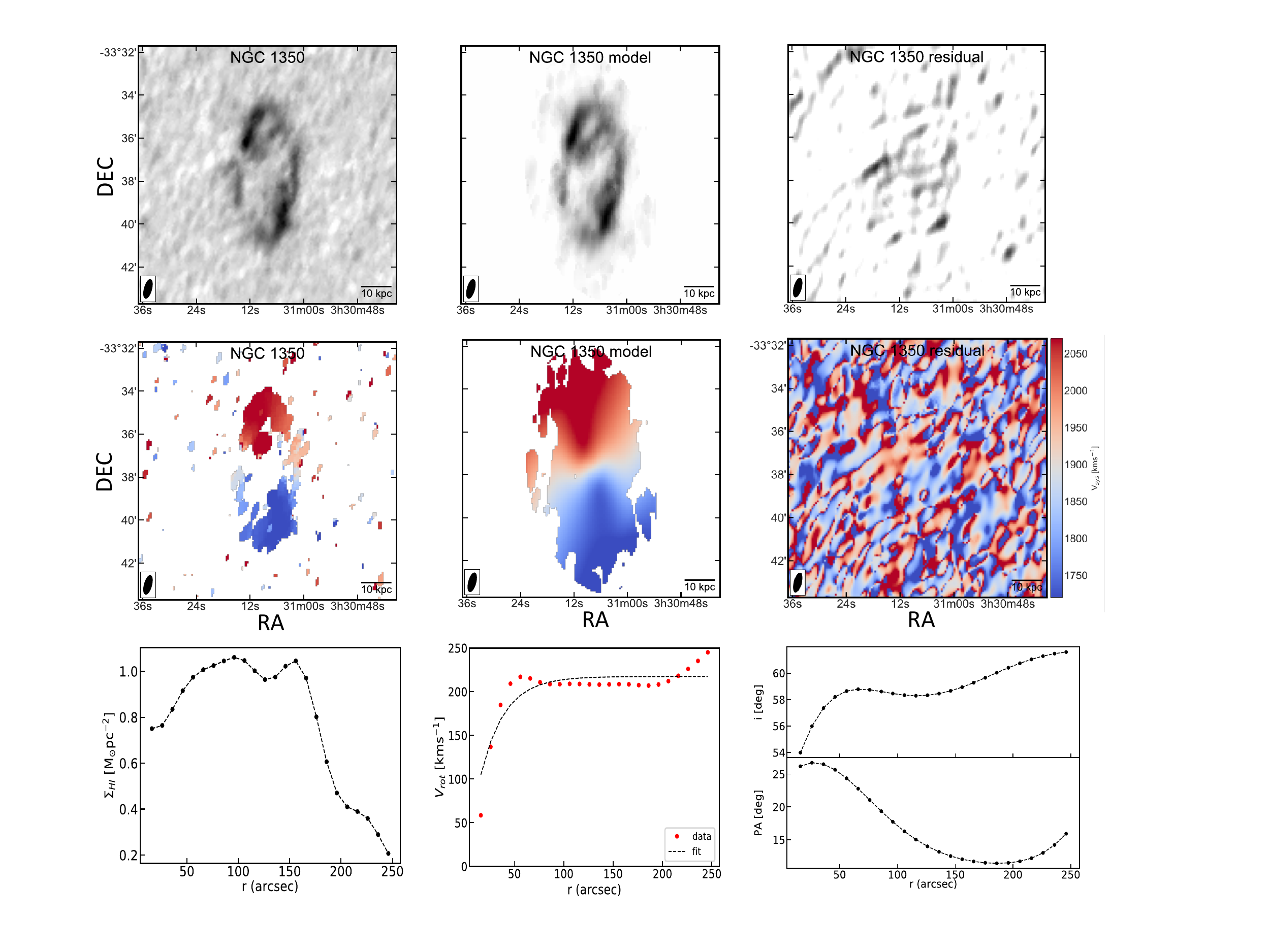}
\caption{NGC 1350: First row shows the moment 0 (intensity) maps of the data, model and the residual. Similarly for the moment 1 (velocity) maps in the second row. The third row shows plots of the \h1 surface density, rotation curve and the fit values for the ring inclination and position angles, derived from the 3D fitting procedure using 3DBarolo.}
\end{figure*}

\begin{figure*}
\hspace*{-1.4cm}
\includegraphics[width=20cm,height=16cm]{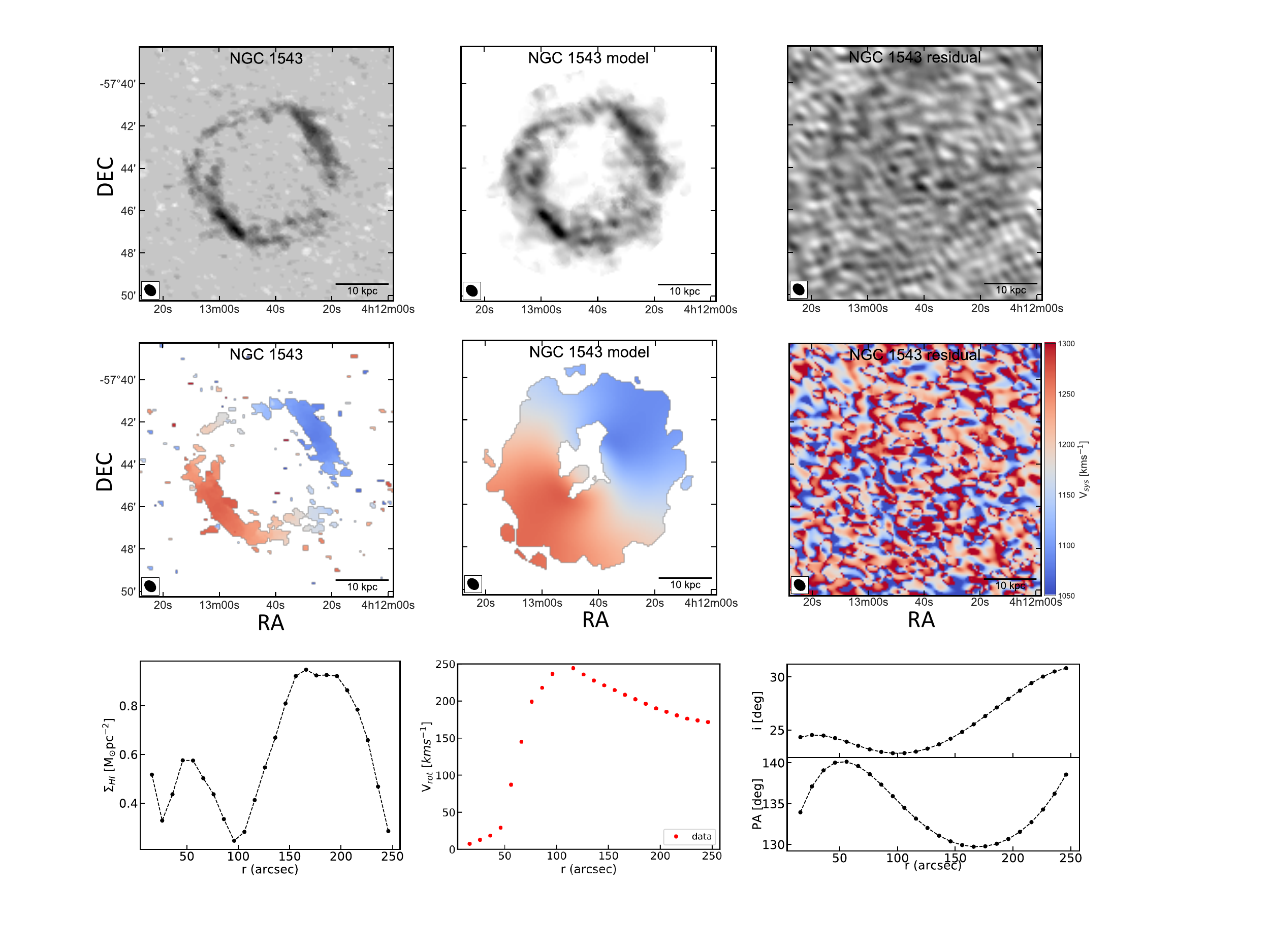}
\caption{NGC 1543: First row shows the moment 0 (intensity) maps of the data, model and the residual. Similarly for the moment 1 (velocity) maps in the second row. The third row shows plots of the \h1 surface density, rotation curve and the fit values for the ring inclination and position angles, derived from the 3D fitting procedure using 3DBarolo. We note that we were unable to make an analytic fit to the rotation curve for this galaxy since it possesses a ring and not a disc. The low S/N in the inner parts of this galaxy may lead to unreliable rotation velocity values derived from the 3D fitting.}
\end{figure*}

\begin{figure*}
\hspace*{-1.4cm}
\includegraphics[width=20cm,height=16cm]{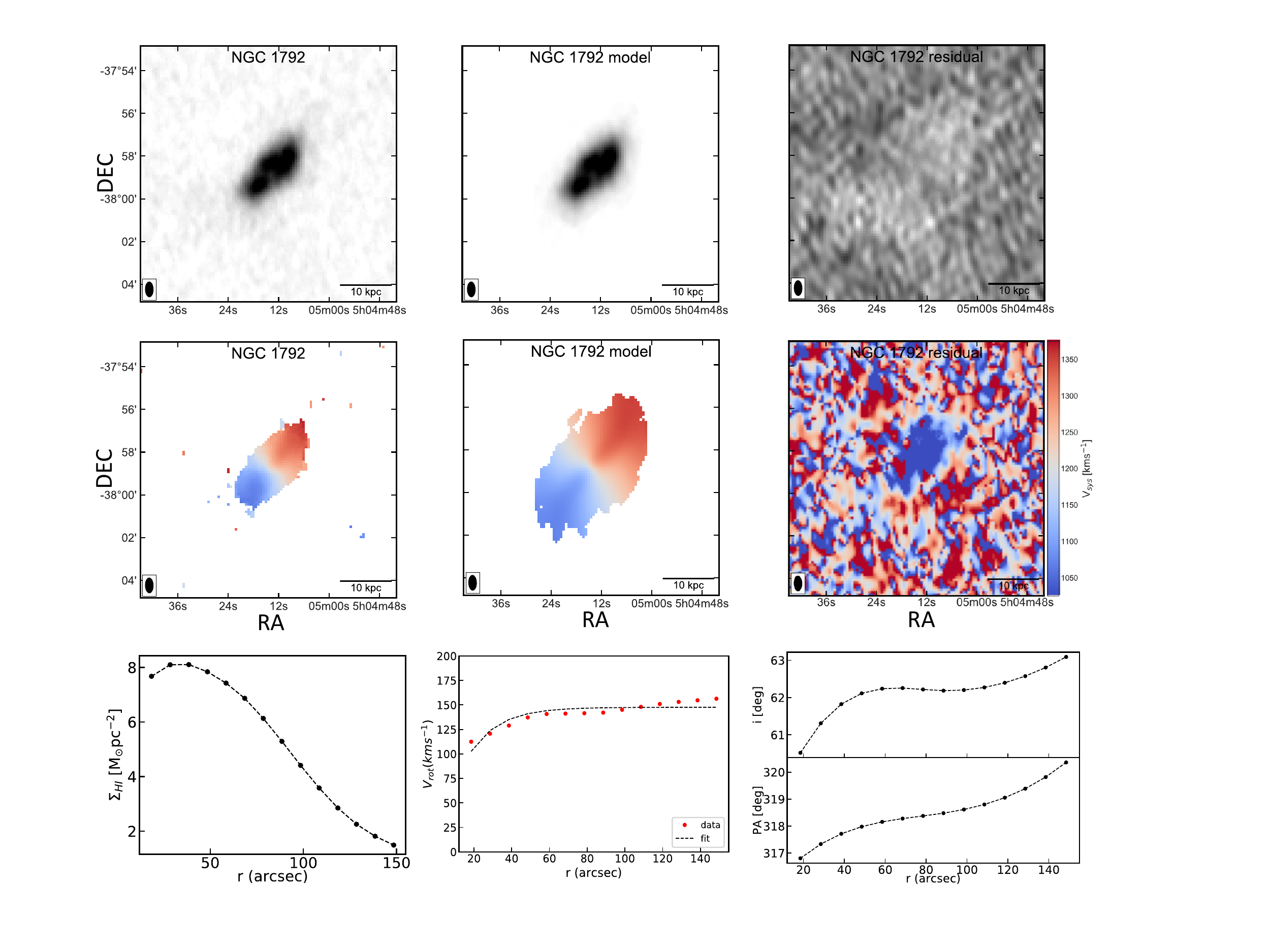}
\caption{NGC 1792: First row shows the moment 0 (intensity) maps of the data, model and the residual. Similarly for the moment 1 (velocity) maps in the second row. The third row shows plots of the \h1 surface density, rotation curve and the fit values for the ring inclination and position angles, derived from the 3D fitting procedure using 3DBarolo.}
\end{figure*}

\begin{figure*}
\hspace*{-1.4cm}
\includegraphics[width=20cm,height=16cm]{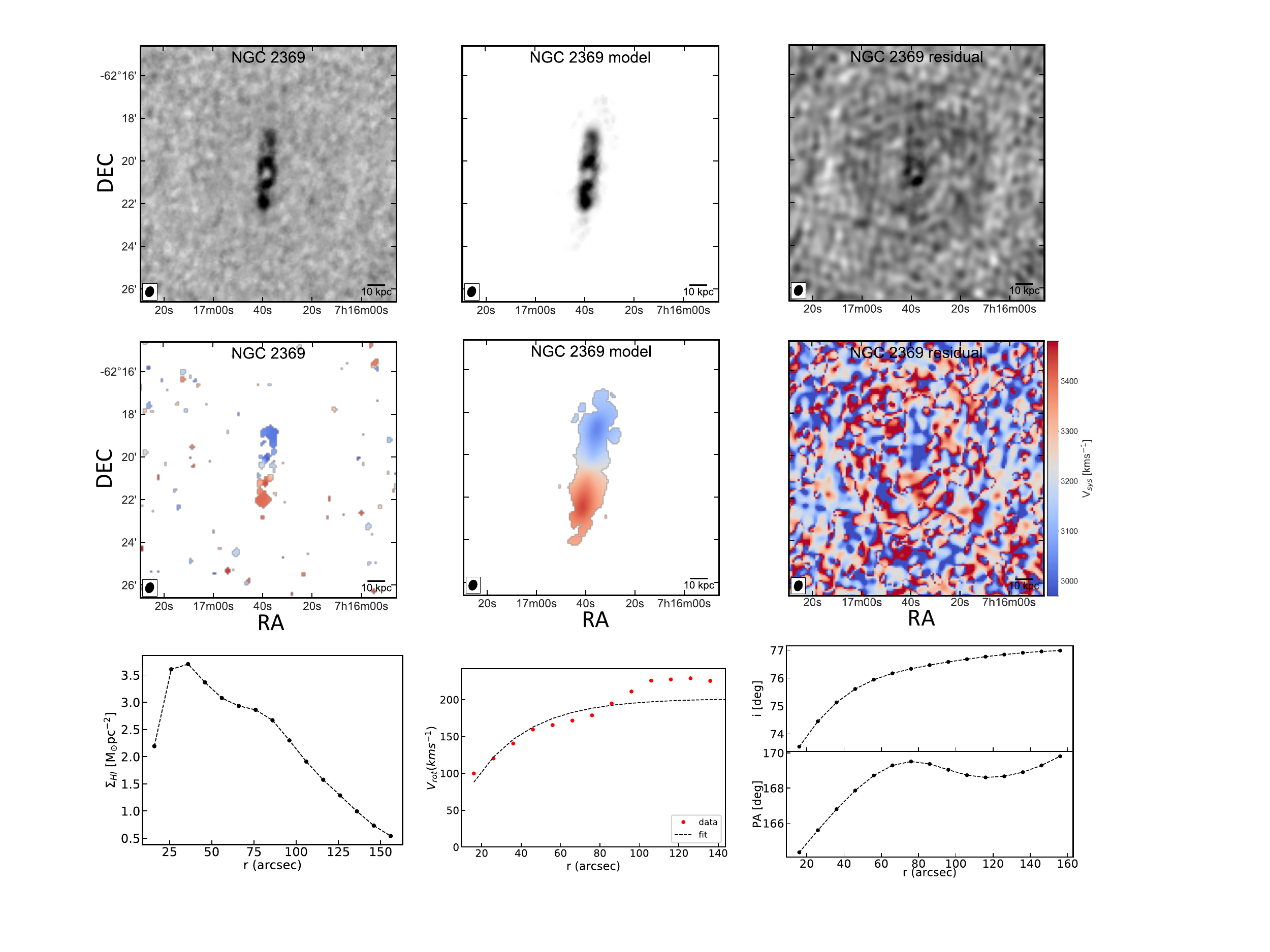}
\caption{NGC 2369: First row shows the moment 0 (intensity) maps of the data, model and the residual. Similarly for the moment 1 (velocity) maps in the second row. The third row shows plots of the \h1 surface density, rotation curve and the fit values for the ring inclination and position angles, derived from the 3D fitting procedure using 3DBarolo.}
\end{figure*}

\begin{figure*}
\hspace*{-1.4cm}
\includegraphics[width=20cm,height=16cm]{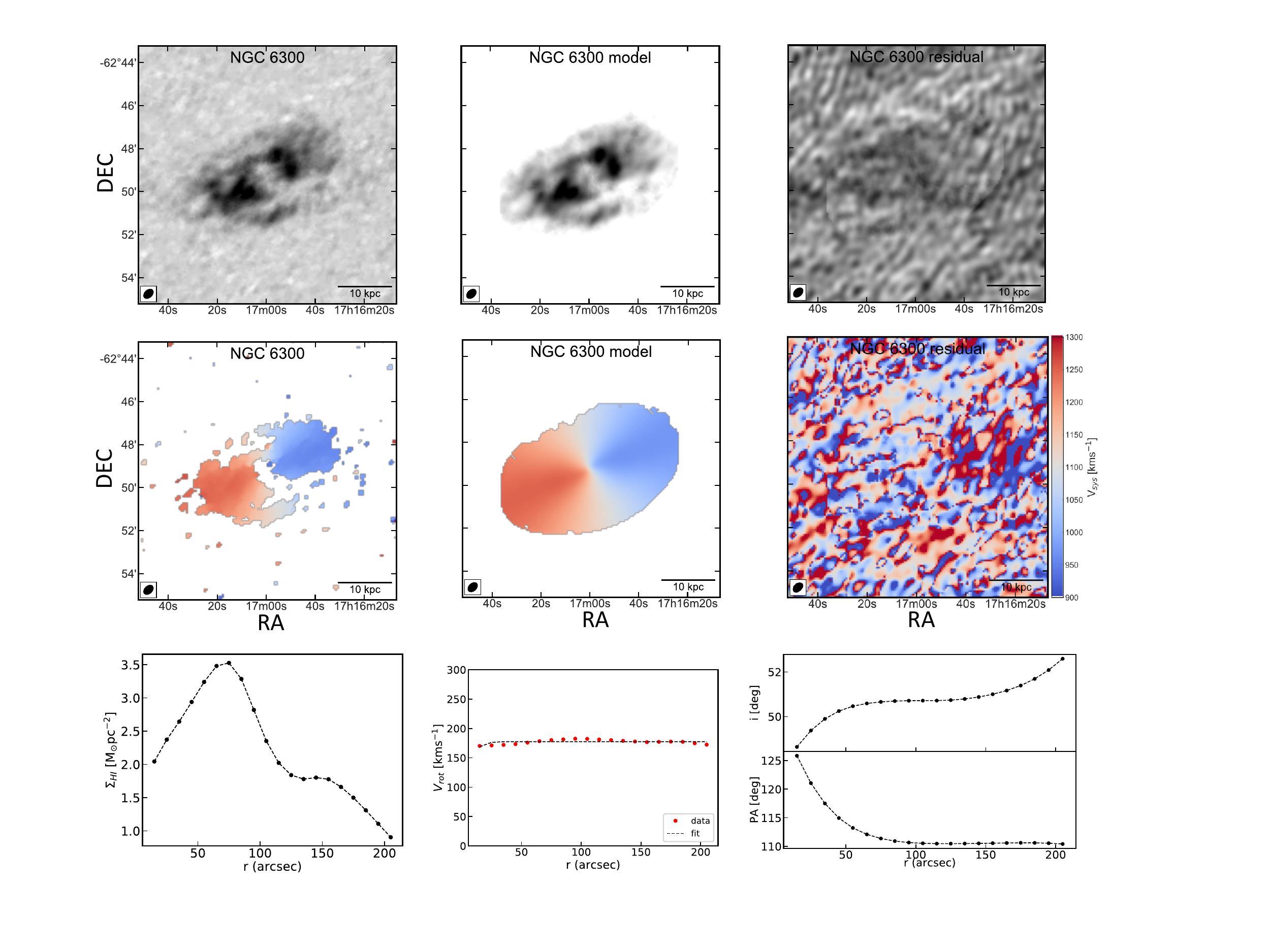}
\caption{NGC 6300: First row shows the moment 0 (intensity) maps of the data, model and the residual. Similarly for the moment 1 (velocity) maps in the second row. The third row shows plots of the \h1 surface density, rotation curve and the fit values for the ring inclination and position angles, derived from the 3D fitting procedure using 3DBarolo.}
\end{figure*}

\begin{figure*}
\hspace*{-1.4cm}
\includegraphics[width=20cm,height=16cm]{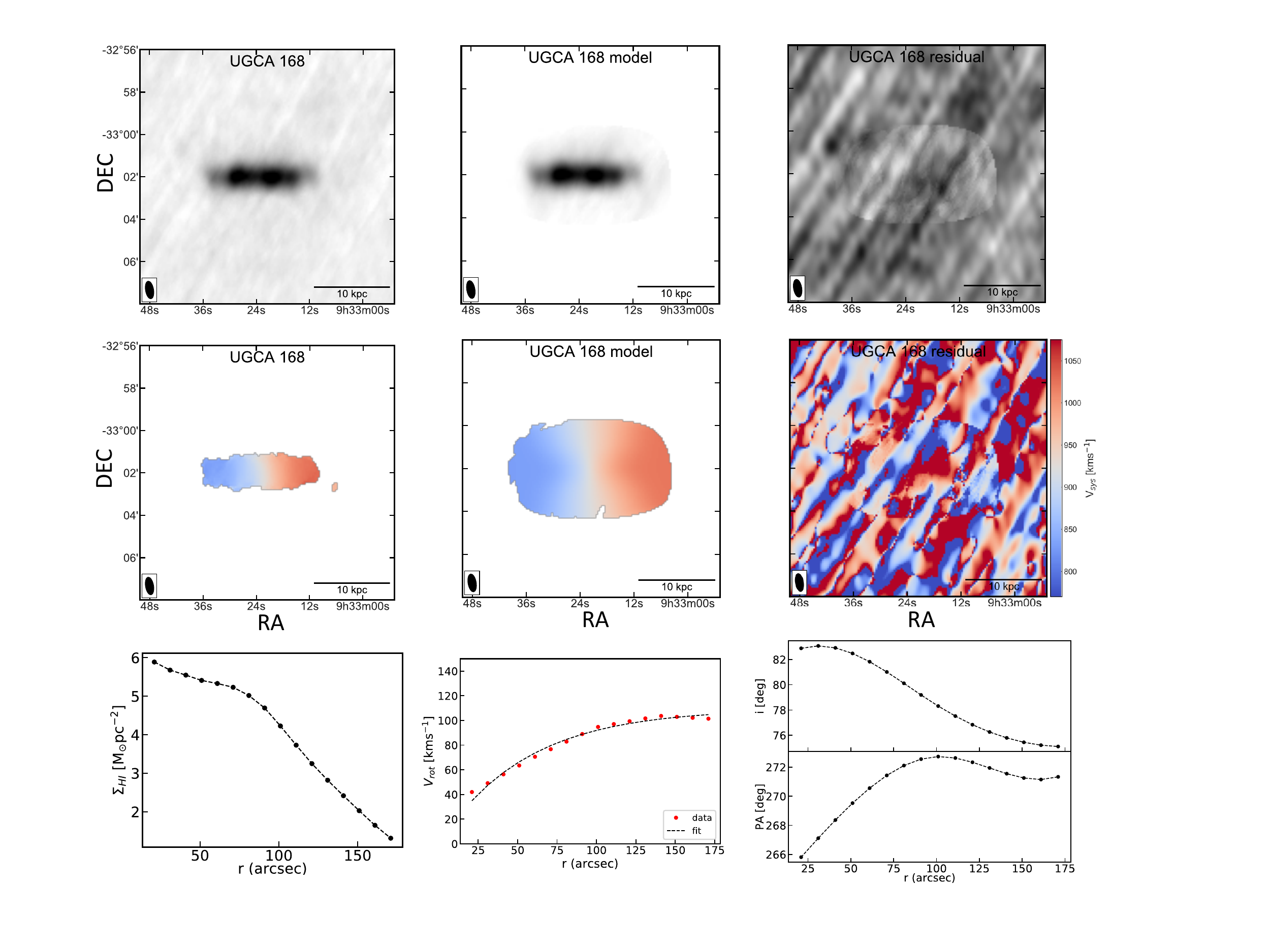}
\caption{UGCA 168: First row shows the moment 0 (intensity) maps of the data, model and the residual. Similarly for the moment 1 (velocity) maps in the second row. The third row shows plots of the \h1 surface density, rotation curve and the fit values for the ring inclination and position angles, derived from the 3D fitting procedure using 3DBarolo.}
\end{figure*}

\bsp	
\label{lastpage}
\end{document}